\documentclass[custom,twocolumn]{webofc_modfd}
\usepackage[varg]{txfonts}   
\usepackage{bbm}
\usepackage{slashed}
\usepackage[usenames,dvipsnames,svgnames]{xcolor}
\usepackage{hyperref}
\hypersetup{
  pdftitle={Monte Carlo event generators and the top quark
    forward--backward asymmetry},
  pdfauthor={J.Winter, P.Z.Skands, B.R.Webber},
  pdfkeywords={Hadron Colliders} {Top Quarks} {Asymmetry},
  colorlinks=true,         
  linkcolor=Blue,    
  citecolor=LimeGreen,     
  filecolor=green,         
  urlcolor=green           
}
\usepackage{hypcap}
\woctitle{Talk given by J Winter @ HCP Symposium 2012, Kyoto, Japan}
\preprint{Cavendish-HEP-13/03\\CERN-PH-TH/2013-026\\MPP-2013-31}
\hcomment{\today}
\begin{document}
\title{Monte Carlo event generators \& the top quark forward--backward
  asymmetry}
%
%
%
\author{%
  Jan~Winter\inst{1,2}\fnsep\thanks{\email{jan.winter@cern.ch}} \and
  Peter~Z.~Skands\inst{2}\fnsep\thanks{\email{peter.skands@cern.ch}} \and
  Bryan~R.~Webber\inst{3}\fnsep\thanks{\email{webber@hep.phy.cam.ac.uk}}
}
\institute{%
  Max-Planck-Institute for Physics, F\"ohringer Ring 6,
  D-80805 Munich, Germany \and
  PH-TH Department, CERN, CH-1211 Geneva 23, Switzerland \and
  Cavendish Laboratory, University of Cambridge, JJ Thomson Avenue,
  Cambridge CB3 0HE, UK
}
\abstract{%
  The leading-order accurate description of $t\bar t$\/ production, as
  usually employed in standard Monte Carlo event generators, gives no
  rise to the generation of a forward--backward asymmetry,
  $A_\mathrm{FB}$. Yet, non-negligible -- differential as well as
  inclusive -- asymmetries may be produced if coherent parton
  showering is used in the hadroproduction of top quark pairs. In this
  contribution we summarize the outcome of our study
  \cite{Skands:2012mm} of this effect. We present a short comparison
  of different parton shower implementations and briefly comment on
  the phenomenology of the colour coherence effect at the Tevatron.
}
\maketitle

\section{Introduction}\label{sec-intro}

\begin{figure}
  \centering\capstart
  \includegraphics[width=0.87\columnwidth,clip]{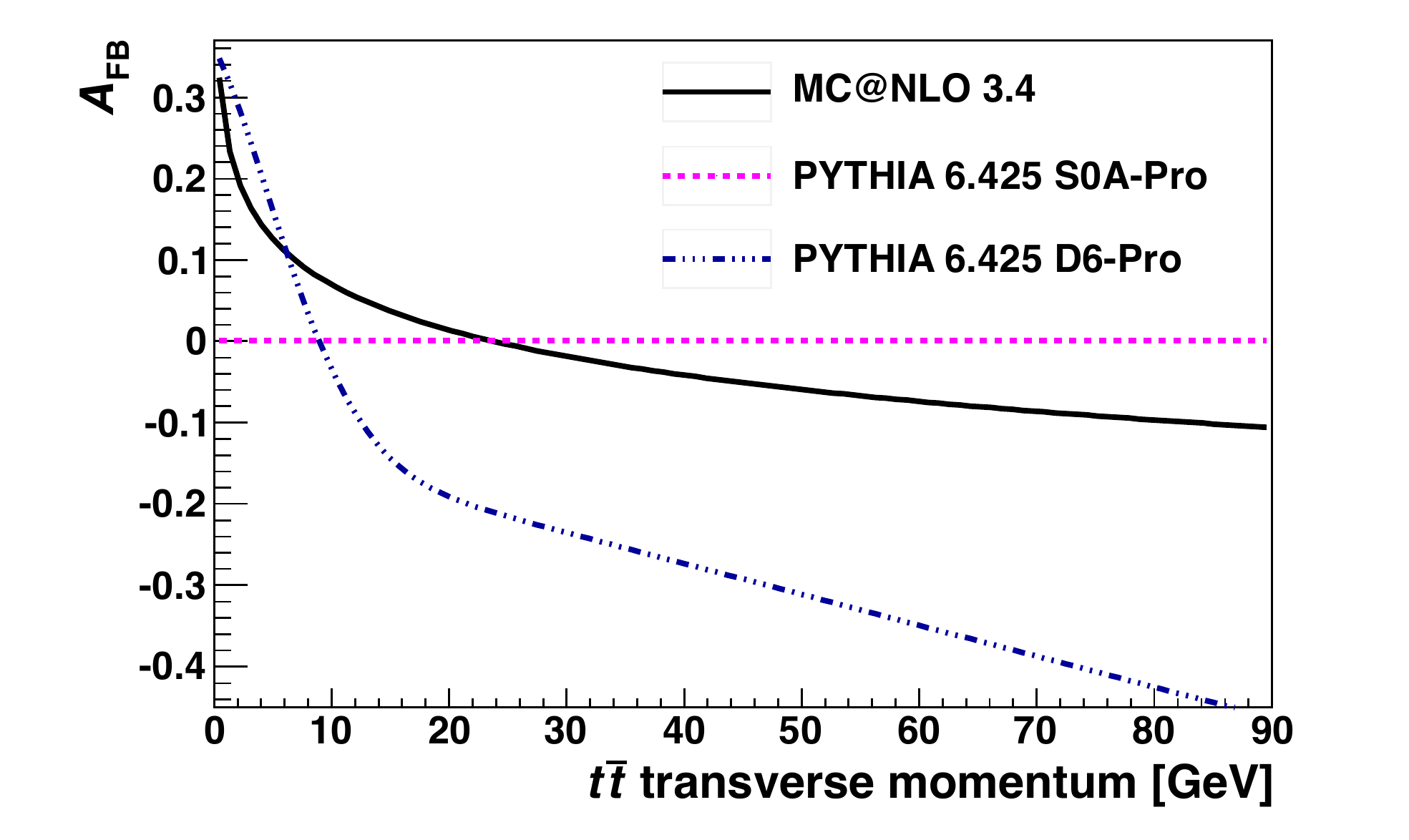}
  \includegraphics[width=0.767\columnwidth,clip]{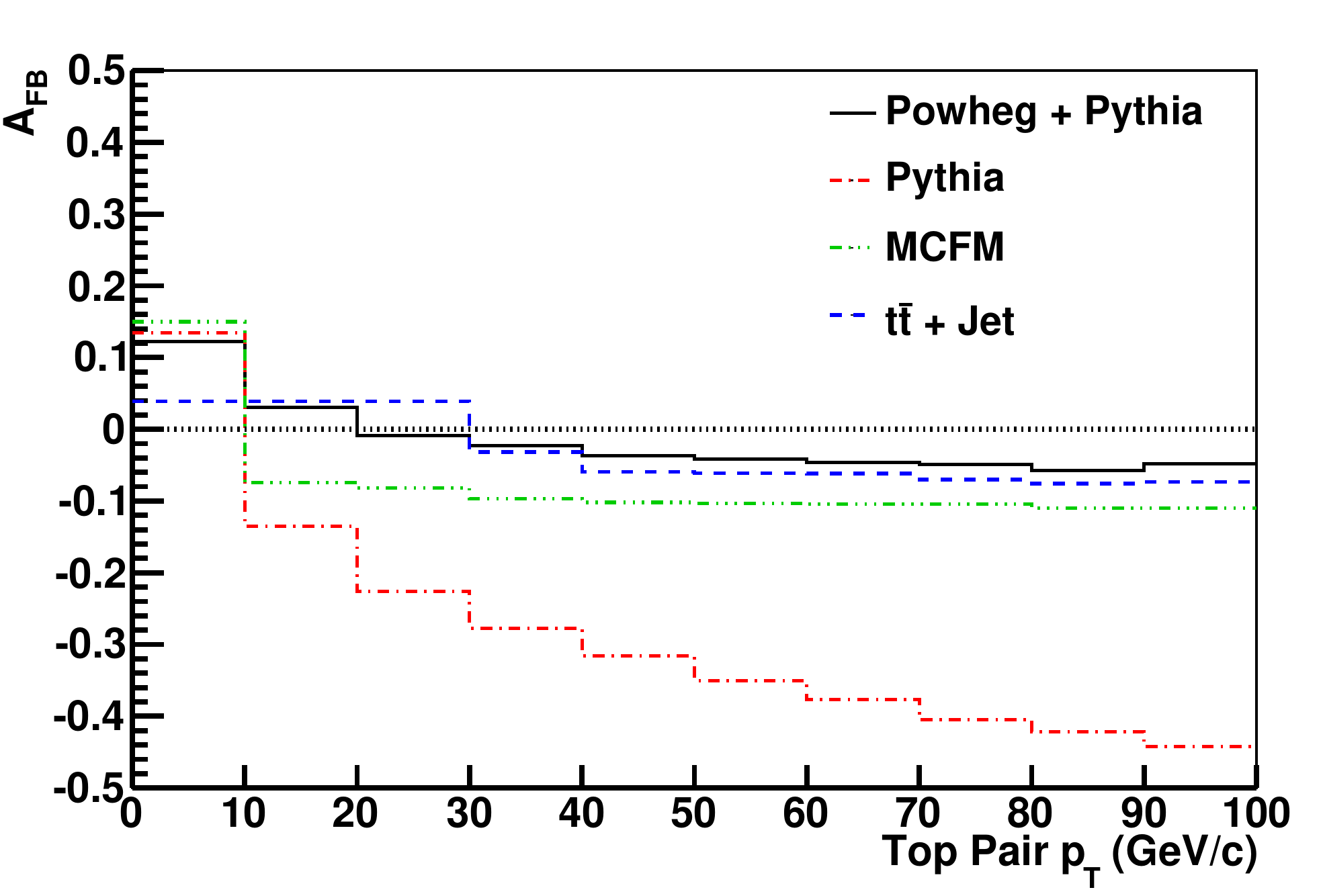}
  \caption{The top quark forward--backward asymmetry $A_\mathrm{FB}$
    in dependence on the transverse momentum of the pair. Results from
    \textsc{MC@nlo}~\cite{Frixione:2003ei},
    \textsc{Pythia}~\cite{Sjostrand:2006za},
    \textsc{Powheg}~\cite{Frixione:2007nw},
    \textsc{MCfm}~\cite{Campbell:2012uf} and
    $t\bar t$+jet production at NLO \cite{Melnikov:2011qx} are shown
    as reported by the D\O\ and CDF Collaborations in
    Refs.~\cite{Abazov:2011rq} and \cite{Aaltonen:2012it}, respectively.}
\label{fig-pTttexp}
\end{figure}

In a number of different measurements the Tevatron Collaborations both
CDF and D\O\ reported results on the top quark forward--backward
asymmetry that clearly lie beyond the Standard Model expectations for
this quantity, see
Refs.~\cite{Aaltonen:2011kc,Abazov:2011rq,Abazov:2012bfa,Aaltonen:2012it}.
They did not only determine inclusive asymmetries but presented also
differential distributions, such as $A_\mathrm{FB}(m_{t\bar t})$,
showing the asymmetry in different bins of observables $O$\/ that
reflect the kinematic properties of the reconstructed top quark
system. One such observable is the transverse momentum of the pair,
$p_{T,t\bar t}$, for which D\O\ made an interesting observation,
documented in Ref.~\cite{Abazov:2011rq}, while studying various Monte
Carlo (MC) predictions: the leading-order (LO) event generator
\textsc{Pythia}~\cite{Sjostrand:2006za}, using the option for
approximate angular coherence, displays a qualitatively similar
$p_{T,t\bar t}$ dependent asymmetry to that of the next-to LO (NLO)
matched parton shower, \textsc{MC@nlo}~\cite{Frixione:2003ei}; cf.~the
top panel in Fig.~\ref{fig-pTttexp}. Disabling the coherent shower
option will restore the expected behaviour for \textsc{Pythia} as
indicated by the dashed line (magenta) in the figure. CDF later on
published similar results~\cite{Aaltonen:2012it} confirming D\O's
observation; this is also shown in Fig.~\ref{fig-pTttexp}, at the
bottom.

The definition of $A_\mathrm{FB}$ used by the Tevatron
experimentalists is based on taking the rapidity difference between
the top and the antitop quark, $\Delta y=y_t-y_{\bar t}$, and dividing
the events according to their $\Delta y$\/ hemisphere:
\begin{equation}\label{eq-gendef} 
  A_\mathrm{FB}(O)\;=\;\frac{\left(\frac{d\sigma}{dO}\right)_+-
                           \left(\frac{d\sigma}{dO}\right)_-}
                          {\left(\frac{d\sigma}{dO}\right)_++
                           \left(\frac{d\sigma}{dO}\right)_-}~.
\end{equation}
The terms $(d\sigma/dO)_\pm$ denote the differential (or, if
$O\equiv\mathbbm{1}$, inclusive) cross sections measured for
forward/backward ($\pm\Delta y>0$) top quark pair configurations; note
that $\Delta y$\/ is a longitudinally boost invariant quantity. We
shall now investigate the effects of colour coherent parton showering
on $A_\mathrm{FB}$ in more detail.

\section{The QCD colour coherence effect}\label{sec-colcoheff}

\begin{figure}
  \centering\capstart
  \includegraphics[width=0.70\columnwidth,clip]{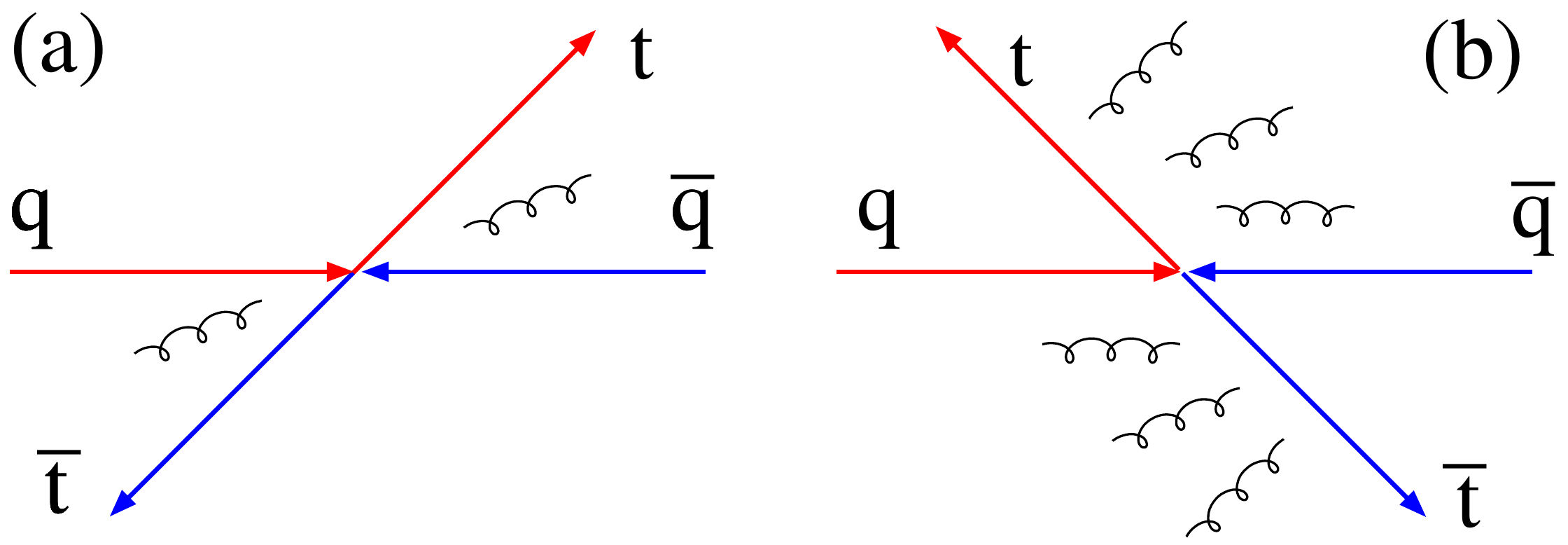}\\[14pt]
  \includegraphics[width=0.85\columnwidth,clip]{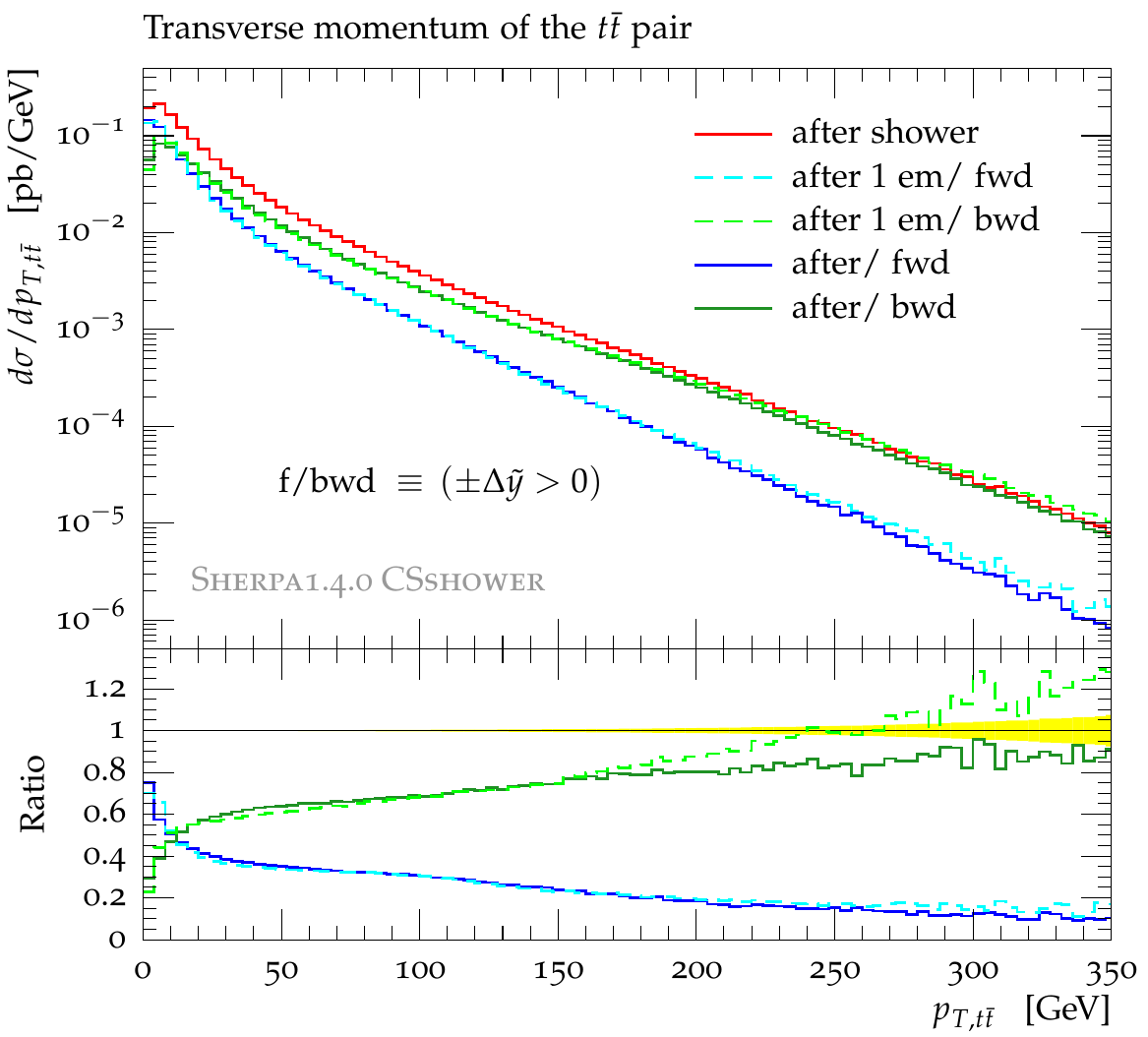}
  \caption{(Top) Colour flow and QCD radiation in (a) forward and (b)
    backward top quark pair production. (Bottom) \textsc{Sherpa}
    \textsc{CSshower} \cite{Schumann:2007mg,Gleisberg:2008ta}
    predictions for all and separately generated forward/backward
    (fwd/bwd) $t\bar t$\/ configurations at LO. The colour coherence
    effect is illustrated after the first and finally all shower
    emissions have occurred by means of the $p_{T,t\bar t}$ distribution.}
  \label{fig-FB}
  \vskip-4mm
\end{figure}

We can gain a qualitative understanding by considering the
large-$N_\mathrm{C}$ colour flows in the leading-order partonic
$q\bar q\to t\bar t$\/ processes, the by far dominant contributions to
$t\bar t$\/ production at the Tevatron. The (leading-)colour flows
associated with these partonic interactions stretch from the initial
to the final state connecting the incoming (anti)quark with the
outgoing (anti)top quark. They form initial--final colour dipoles as
shown in the upper part of Fig.~\ref{fig-FB}. The deflection of the
top quark out of the incoming quark's direction is a measure of the
acceleration of the colour charge. Colour coherence manifests itself
through this deflection angle: a mild scatter characterized by a small
opening angle, keeping the forward motion of the top quark, only induces
weak radiation off this $qt$\/ dipole, see Fig.~\ref{fig-FB} upper
left. In contrast, strong scattering and re-direction of the top quark
leads to a more violent acceleration of colour and hence a larger
number of potentially also harder QCD emissions. Top quark pairs in
backward configurations therefore experience an increased recoil which
we illustrate in the upper right of Fig.~\ref{fig-FB}. Consequently,
we find a larger number of soft $p_{T,t\bar t}$ events correlated with
positive values of the asymmetry while hard $p_{T,t\bar t}$ events
often contribute to negative asymmetries -- as well known from the NLO
calculation, cf.~Fig.~\ref{fig-pTttexp}.

We can easily test this rationale in a MC simulation which we do by
utilizing \textsc{Sherpa}'s \textsc{CSshower}
implementation~\cite{Schumann:2007mg}.\footnote{The use of dipole-like
  splitting kernels derived from spin-averaged, leading-colour reduced
  Catani--Seymour subtraction terms ensures an accurate treatment of
  the soft limit of QCD emission, thus, enables the construction of a
  colour coherent radiation pattern.}
We track the evolution of top quark pairs generated at leading, fixed
order in forward ($\Delta\tilde y>0$) and backward ($\Delta\tilde y<0$)
configurations independently. The tilde sign in $\Delta\tilde y$\/ is
used to flag an evaluation at the matrix-element level. The results in
the lower part of Fig.~\ref{fig-FB} clearly demonstrate that after
the coherent showering phase, initial backward configurations yield a
harder $p_{T,t\bar t}$ spectrum than initially forwards moving top quark
pairs. The latter preferably populate the soft region of
$p_{T,t\bar t}\lesssim10\mathrm{\:GeV}$, and both observations thus
agree well with our qualitative explanation from above.

\subsection{Analytic approximation of the effect}

\begin{figure}
  \centering\capstart
  \includegraphics[width=0.75\columnwidth,clip]{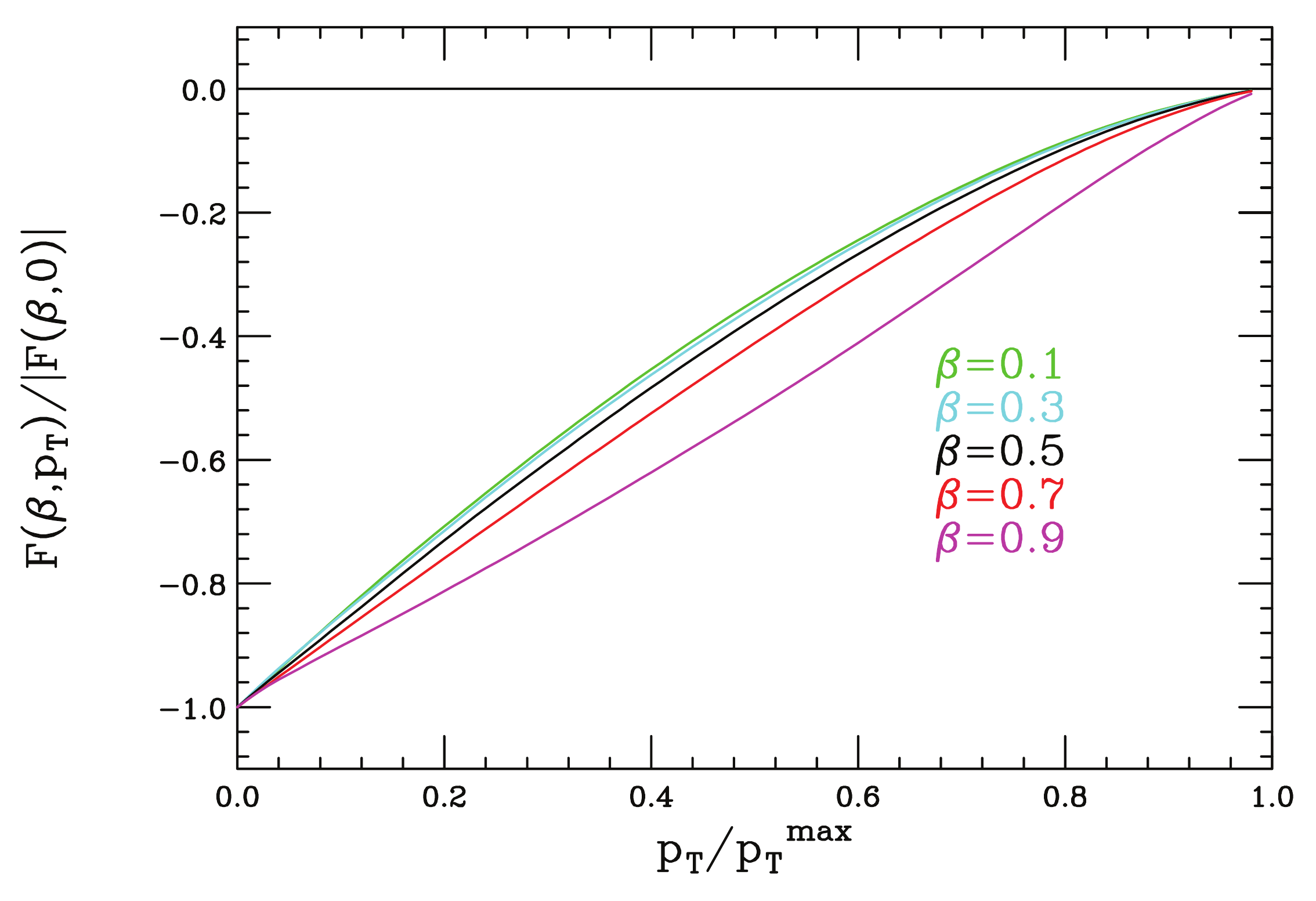}
  \caption{The function $F(\beta,p_T)$, normalized to its absolute
    value at zero $p_T$; the maximum $p_T$ is given as
    $p^\mathrm{max}_T=\beta^2\sqrt{\hat s}/2$.}
  \label{fig-fofbetapT}
\end{figure}

\begin{figure}
  \centering\capstart
  \sidecaption
  \includegraphics[width=0.58\columnwidth,clip]{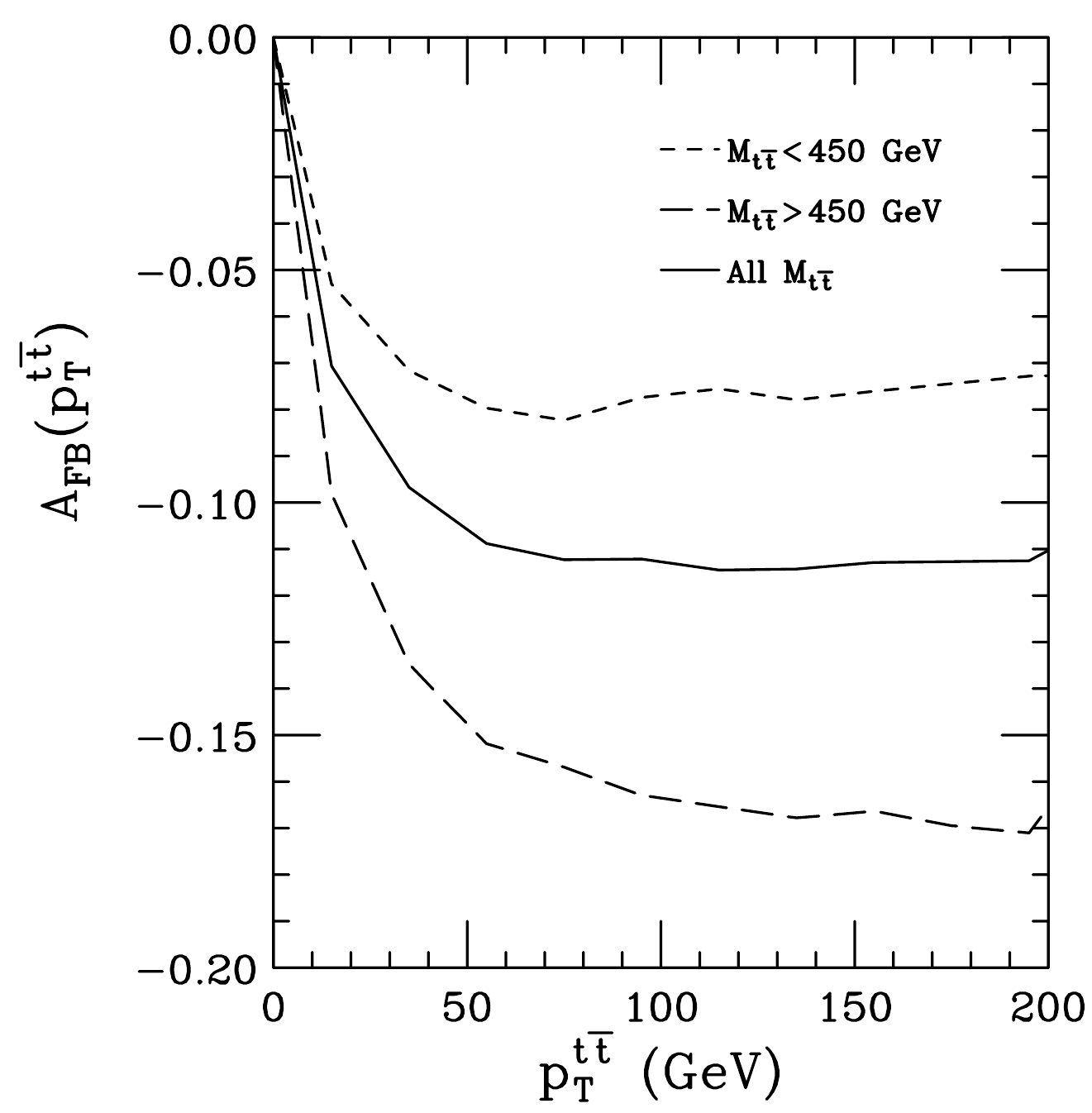}
  \caption{Leading-order QCD predictions (calculated with
    \textsc{MCfm}~\cite{Campbell:2012uf} at NLO in $t\bar t$\/
    production) for the forward--backward asymmetry as a function of
    $p^{t\bar t}_T$ imposing an upper, a lower and no bound at all on
    the $t\bar t$\/ pair mass.}
  \label{fig-mcfm}
\end{figure}

The qualitative picture which we argued for in
Sec.~\ref{sec-colcoheff} can be made more explicit. Analyzing the
real-emission contribution to the asymmetry, mainly arising from
$q\bar q\to t\bar tg$\/ processes, one can show, see
Ref.~\cite{Skands:2012mm}, that the differential dependence of the
asymmetry on the pair transverse momentum can be written as
\begin{equation}\label{eq-ptdep}
  \frac{p_T}{\hat\sigma_\mathrm{B}}\,\frac{d\hat\sigma_\mathrm{A}}{dp_T}\;=\;
  \frac{\alpha_\mathrm{s}}{\pi}\,\frac{N^2_\mathrm{C}-4}{N_\mathrm{C}}\,
  F(\beta,p_T)
\end{equation}
where $p_T\equiv p_{T,t\bar t}$ and $\beta=\sqrt{1-4m_t^2/\hat s}$
denotes the top quark center-of-mass velocity. The Born and asymmetry
cross sections are given by $\hat\sigma_\mathrm{B}$ and
$\hat\sigma_\mathrm{A}$, respectively.

\begin{figure*}
  \centering\capstart
  \includegraphics[width=0.41\textwidth,clip]{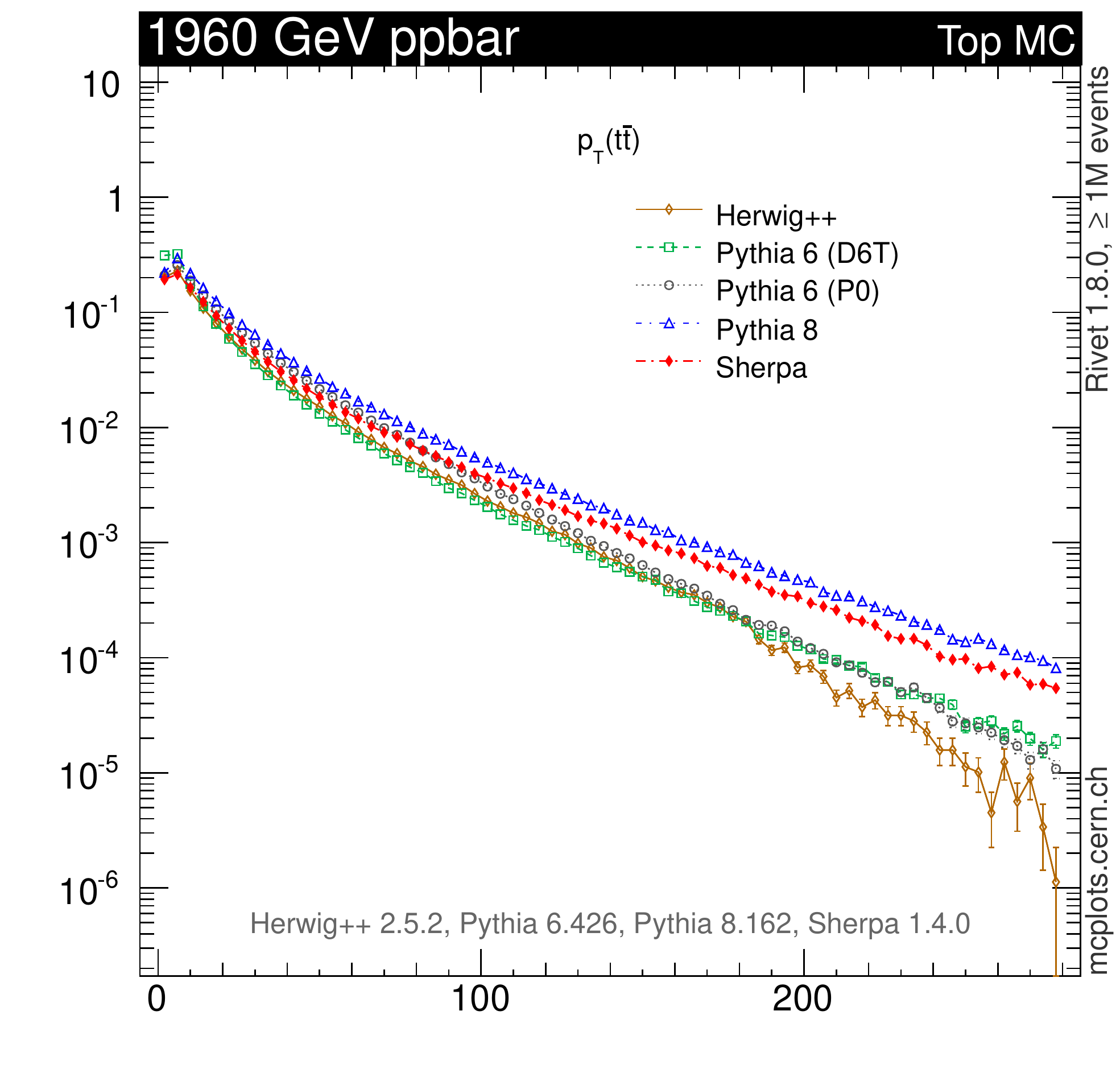}\hskip7mm
  \includegraphics[width=0.41\textwidth,clip]{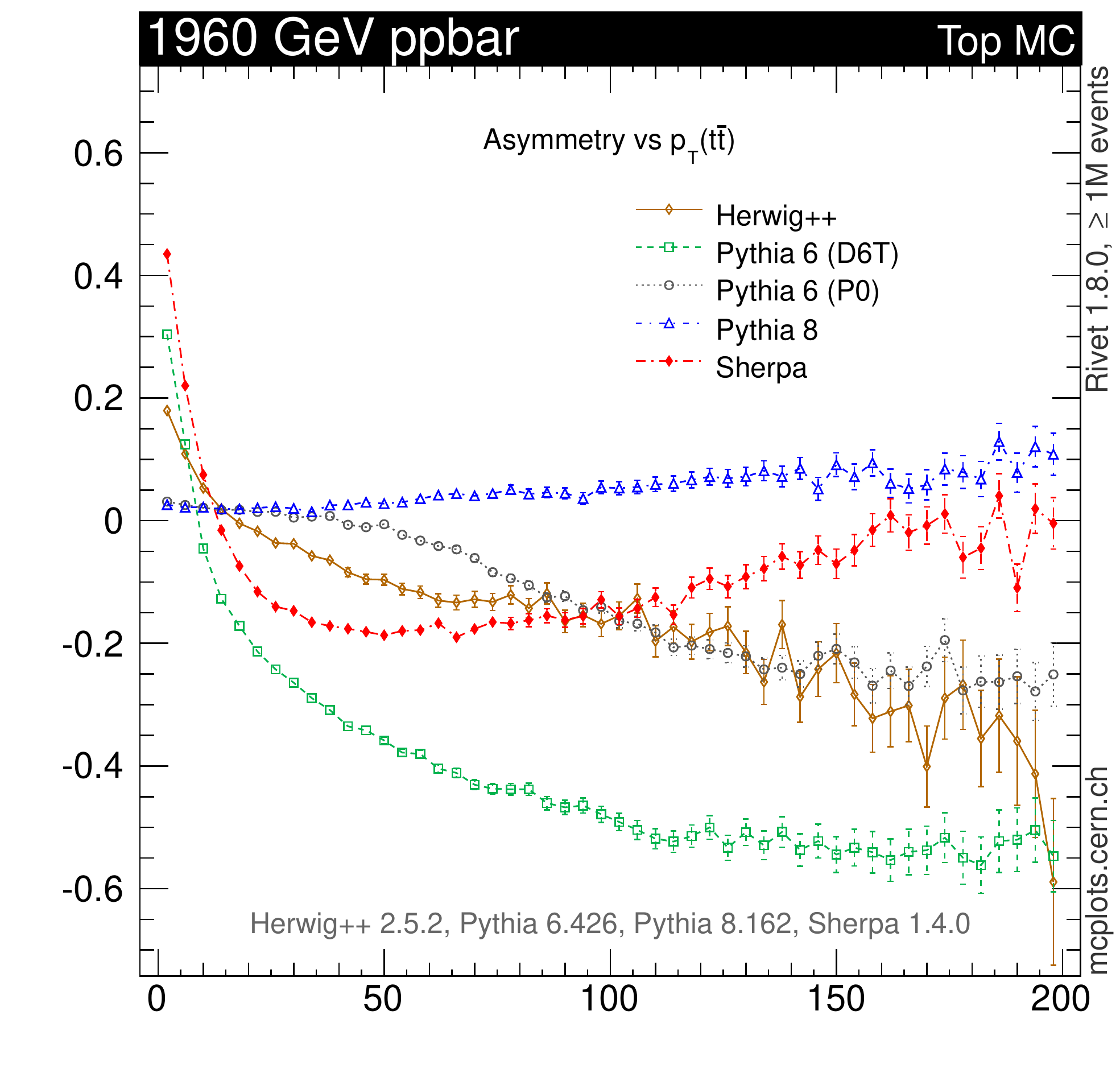}
  \includegraphics[width=0.41\textwidth,clip]{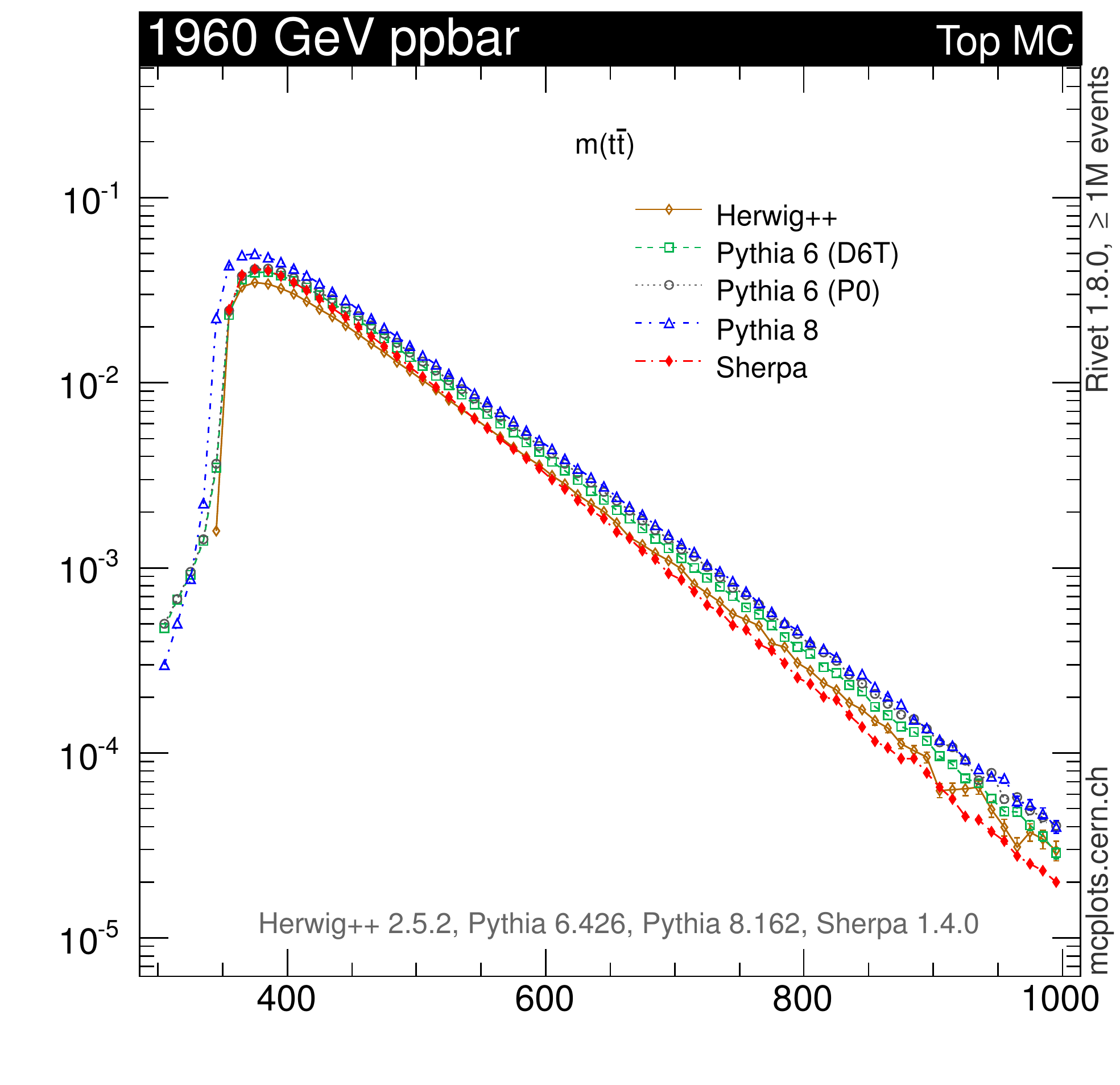}\hskip7mm
  \includegraphics[width=0.41\textwidth,clip]{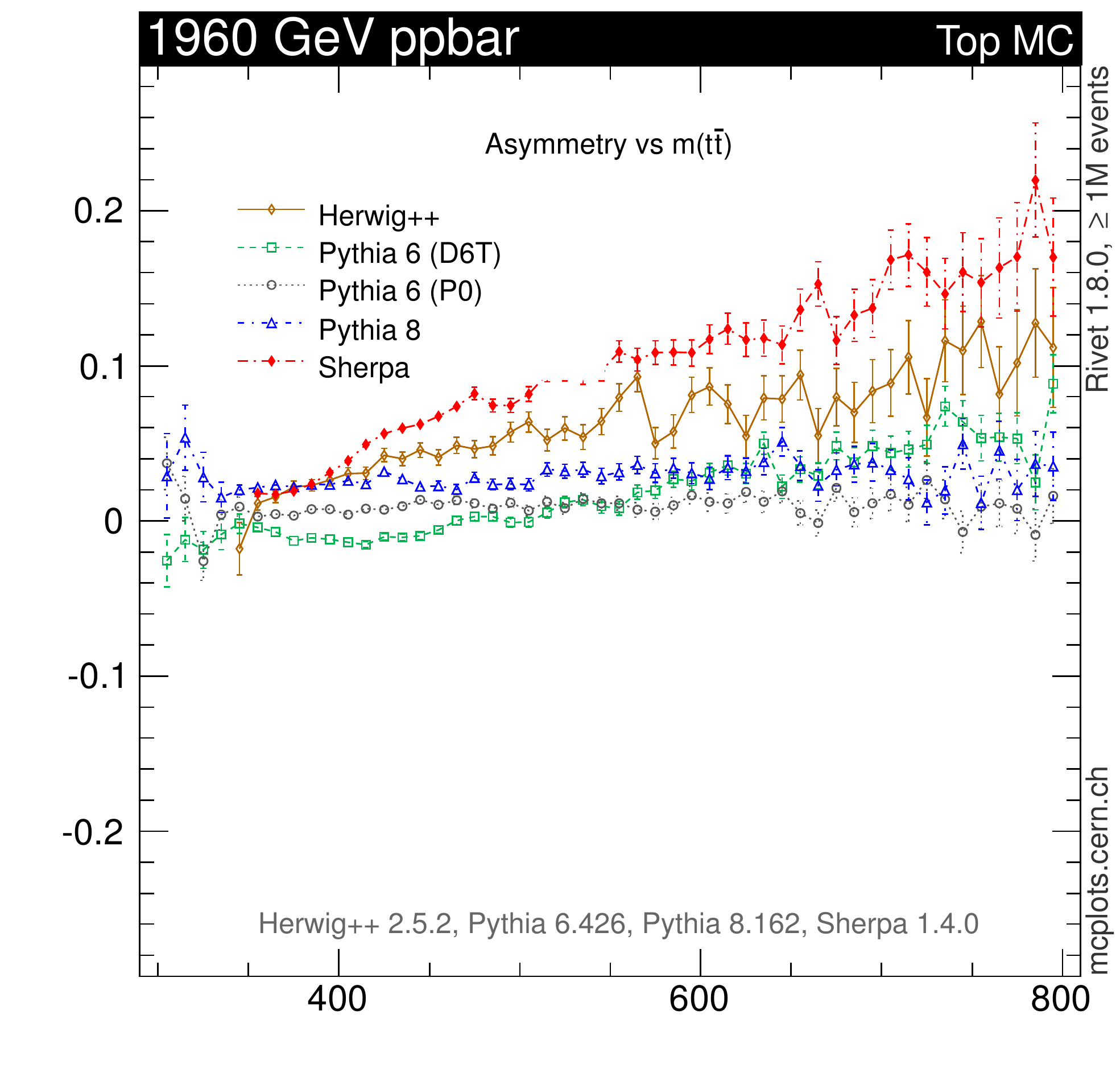}
  \caption{Differential cross sections $d\sigma/dO$\/ (in
    $\mathrm{pb/GeV}$, left panels) and asymmetries $A_\mathrm{FB}(O)$
    (right panels) as a function of the $t\bar t$\/ pair transverse
    momentum ($p_{T,t\bar t}$ in $\mathrm{GeV}$, upper row) and $t\bar
    t$\/ pair mass ($m_{t\bar t}$ in $\mathrm{GeV}$, lower
    row). Various predictions (and their MC errors) given by LO event
    generators are compared to each other.}
  \label{fig-pTtt-mtt}
  \vskip-2mm
\end{figure*}

We can now discuss to what extent a coherent branching algorithm
reproduces this functional form given in Eq.~(\ref{eq-ptdep}). Such
shower generators treat the gluon radiation as coherent emission from
the top quarks lines of the Born process, therefore overestimate the
colour factor in Eq.~(\ref{eq-ptdep}) by
$2\,C_\mathrm{F}=(N^2_\mathrm{C}-1)/N_\mathrm{C}$ (a 60\%
overestimate) and approximate the kinematic part of the asymmetry
amplitude by the corresponding Born term times dipole-like/eikonal
factors. That means we effectively obtain a description of the
asymmetry which is correct in the soft gluon limit (apart from the
colour factor), but only approximate for $p_T>0$, i.e.\ in the
coherent shower generators the coupling- and colour-stripped asymmetry
function $F(\beta,p_T)$ given in Eq.~(\ref{eq-ptdep}) can only be
``underestimated'' by $F(\beta,0)$ which reads
$F(\beta,0)=-4\,\beta-\beta^3+{\cal O}(\beta^5)$ when Taylor expanded
in $\beta$. Note that $F(\beta,p_T)$ becomes less negative the further
away from the $p_T=0$ limit. This is depicted in
Fig.~\ref{fig-fofbetapT} for different fixed values of $\beta$\/
where we observe larger deviations for increasing top quark pair
invariant masses. Both the colour factor and kinematic approximations
lead to a more pronounced $A_\mathrm{FB}(p_{T,t\bar t})$ dependence, as
seen in Fig.~\ref{fig-pTttexp} when comparing the \textsc{Pythia}
coherent shower prediction with that given by \textsc{MC@nlo}.

Finally, Fig.~\ref{fig-mcfm} shows the $p_T>0$ dependence of the
asymmetry as defined in Eq.~(\ref{eq-gendef}) which one obtains from a
full NLO calculation for $t\bar t$\/ production at the
Tevatron.\footnote{The finite part of the virtual correction leads to
  a positive delta peak contribution at $p_T\equiv0$ which however is
  not shown here.}
It is interesting to note that the asymmetry approaches zero but does
not become positive as $p_T\to0$. This is because the singularity
structures present in the numerator and denominator are different. The
denominator diverges faster owing to initial-state collinear
singularities that cancel in the numerator. In contrast to the
behaviour at fixed order, NLO matched or coherent parton showers
predict an $A_\mathrm{FB}(p_{T,t\bar t})$ cross-over, as shown in
Fig.~\ref{fig-pTttexp}, at low but non-negligible values of $p_T$.
This is a consequence of the Sudakov suppression of small $p_T$
occurring due to multiple soft gluon emission, an effect absent in the
fixed-order description. This Sudakov suppression yields a spreading
of the positive asymmetry over a finite region of $p_T\gtrsim0$.

\section{Differential asymmetry predictions}\label{sec-dafb}

Considering the impact colour coherence has on generating differential
asymmetries, it is important to study the response of standard MC
event generators using $t\bar t$\/ production (without decays) at the
Tevatron. For a representative collection of parton showers, we
present their respective asymmetry predictions in
Fig.~\ref{fig-pTtt-mtt} together with the associated differential
cross sections as functions of $p_{T,t\bar t}$ (upper row) and
$m_{t\bar t}$ (lower row). Far more details regarding this MC
tools comparison and the choice of parameters can be found in
Ref.~\cite{Skands:2012mm}. Here we only highlight a small fraction of
the results of this comparison.

As one can see from Fig.~\ref{fig-pTtt-mtt}, default
\textsc{Herwig++}~\cite{Bahr:2008pv} and \textsc{Sherpa}~\cite{Gleisberg:2008ta}
produce differential $A_\mathrm{FB}$ sufficiently similar to what one
expects from the approximate NLO treatment mentioned above, especially
the rise of $A_\mathrm{FB}$ with increasing $m_{t\bar t}$ is remarkable.
Using the $m_{t\bar t}$ observable the Sudakov region is stretched/applied
over the entire mass range which leads to a (an almost entirely)
positive $A_\mathrm{FB}$ dependence. Both \textsc{Herwig++} (through
angular ordering) and \textsc{Sherpa} (through dipole showering) have
implemented QCD coherence in a proper way, however rely on different
strategies regarding initial shower conditions and treatment of
recoils. This induces differences between their predictions as seen in
Fig.~\ref{fig-pTtt-mtt}. While the version of
\textsc{Pythia~8}~\cite{Sjostrand:2007gs} used here does not yet
account for QCD coherence effects, hence produces rather small,
$p_{T,t\bar t}$ as well as $m_{t\bar t}$ insensitive asymmetries, the
\textsc{Pythia~6}~\cite{Sjostrand:2006za} predictions in
Fig.~\ref{fig-pTtt-mtt} are examples for the fact that the older
generation provides options (tunes) with varying amount of coherence.
The D6T and P(erugia)0 tunes~\cite{Skands:2010ak} emerged from efforts
to improve the description of Tevatron and LHC data, respectively. In
particular the D6T tune amplifies the effect of colour coherence as
explained in Ref.~\cite{Skands:2012mm}.

\section{Inclusive asymmetry and longitudinal recoil effects}\label{sec-iafb}

\begin{figure}[t!]
  \centering\capstart
  \includegraphics[width=0.85\columnwidth,clip]{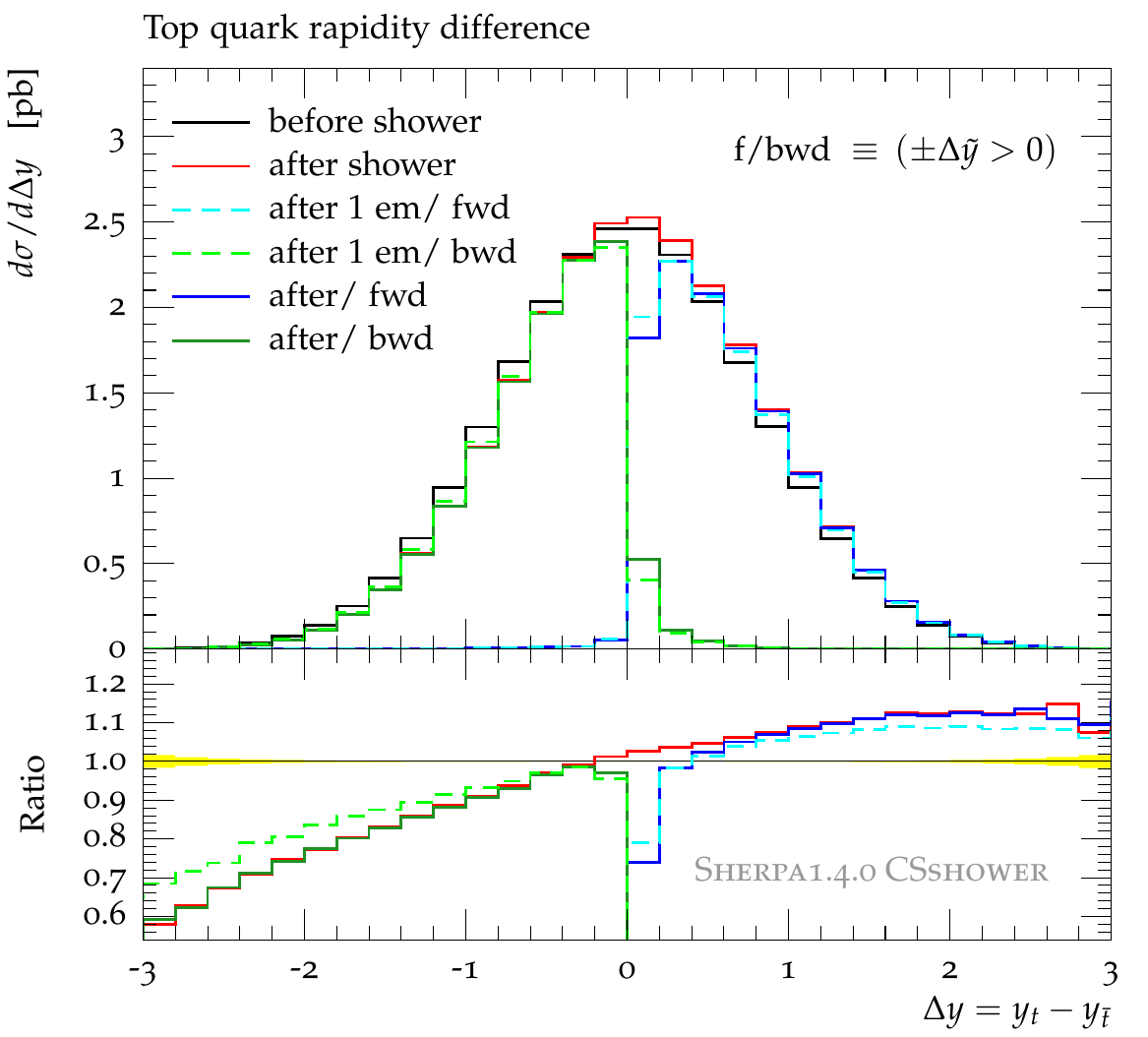}
  \caption{The $\Delta y$\/ distributions for various modes of
    \textsc{CSshower}-ing applied after disjoined as well as joined
    $\Delta\tilde y$\/ hemisphere generation at leading, fixed order
    in top quark pair production.}
  \label{fig-migrate}
\end{figure}

Inspecting the numbers in Tab.~\ref{tab-1}, it is obvious that colour
coherent showering not only produces non-trivial differential
asymmetries but also an inclusive $A_\mathrm{FB}$.\footnote{This can
  also be anticipated by viewing \textsc{Herwig++}'s and
  \textsc{Sherpa}'s positive-valued $A_\mathrm{FB}(m_{t\bar t})$
  results given in Fig.~\ref{fig-pTtt-mtt}.}
This clearly comes as another surprise, but can be explained via event
migrations arising from longitudinal recoil effects. This is specified
in Ref.~\cite{Skands:2012mm} where we also show that the effect,
already at LO, may give a relevant contribution to the inclusive
asymmetry. Here, we only want to provide evidence that migration
indeed occurs in coherent showers, using the \textsc{CSshower} results
displayed in Fig.~\ref{fig-migrate}.

\begin{table}[t!]
  \centering
  \caption{Inclusive $A_\mathrm{FB}$ (in \%) in each shower model, and
    to leading, non-trivial order in QCD (taken from \textsc{MCfm}).}
  \label{tab-1} 
  \small
  \begin{tabular}{lcclcc}\hline\\[-3mm]
    Model                     & Version & $A_\mathrm{FB}$ &
    Model                     & Version & $A_\mathrm{FB}$\\\hline\\[-3mm]
    \textsc{Pythia~6}         & 6.426   & $-0.1$ &
    \textsc{Sherpa}           & 1.4.0   &  $5.5$\\
    \textsc{Pythia~8}         & 8.163   &  $2.5$ &
    QCD                       & LO      &  $6.0$\\
    \textsc{Herwig++}\!\!\!\! & 2.5.2   &  $3.9$\\\hline
  \end{tabular}
\end{table}

Following the same strategy used to produce the $p_{T,t\bar t}$
spectra presented in Fig.~\ref{fig-FB}, it is straightforward to
obtain the corresponding $\Delta y$\/ distributions to examine the
migration effect directly. A simple visualization of the longitudinal
recoil effect comes in terms of a stretching/widening of the gluon
emitting $qt$\/ initial--final dipole. Independently of the magnitude
of the initial scattering angle, the top quark will be slightly pushed
forward while the antitop quark retains its direction such that
$\Delta y=\Delta\tilde y+\epsilon$\/ is a plausible parametrization
($\epsilon>0$). The results shown in Fig.~\ref{fig-migrate} confirm
this conjecture. Several observations are made: (1) the imbalance
between $\Delta y<0$ and $\Delta y>0$ events, i.e.\ the generated
asymmetry, is clearly visible by comparing the after- with the
before-shower result (cf.\ red vs.\ black line), (2) the migration of
initial backward and forward configurations to higher values of
$\Delta y$\/ produces an overflow of events of the former category
closing the dip at $\Delta y\gtrsim0$ generated by shifted events of
the latter category (cf.\ green vs.\ blue lines), (3) $+\to-$
migrations in $\Delta y$\/ (i.e.\ $\epsilon<0$) are suppressed and (4)
the largest contribution to the migration effect already appears with
the first emission (cf.\ dashed vs.\ solid lines). Generally,
migrations are small, occur locally and clearly favour $\epsilon>0$.
They therefore fuel the generation of a positive-valued inclusive
asymmetry.

\section{Phenomenological tests}\label{sec-phen}

With our understanding of the origin of the colour coherence effect,
we can study some of its phenomenological implications which we
briefly discuss below.

\subsection{Asymmetry in different phase-space domains}

\begin{table}[t!]
  \centering
  \caption{Inclusive $A_\mathrm{FB}$ results (in \%) imposing simple
    kinematic cuts ($M=450\mathrm{\:GeV}$, $Q=50\mathrm{\:GeV}$) for
    the \textsc{Herwig++} model, the \textsc{CSshower} and $t\bar t$\/
    production at NLO (computed with \textsc{MCfm}).}
  \label{tab-2}\small
  \begin{tabular}{lcrlrl}\hline\\[-3mm]
    Model & Version &
    $m_{t\bar t}<M$ & $>M$ & $p_{T,t\bar t}<Q$ & $>Q$\\\hline\\[-3mm]
    \textsc{Herwig++}\!\!\!\! & 2.5.2 & $2.7$&$6.0$ &$5.8$&$-14.3$\\
    \textsc{Sherpa}           & 1.4.0 & $3.5$&$9.2$ &$8.7$&$-15.4$\\
    QCD                       & LO    & $4.1$&$9.3$ &$7.0$&$-11.1$\\\hline
  \end{tabular}
\end{table}

\begin{figure*}[t!]
  \centering\capstart
  \includegraphics[width=0.33\textwidth,clip]{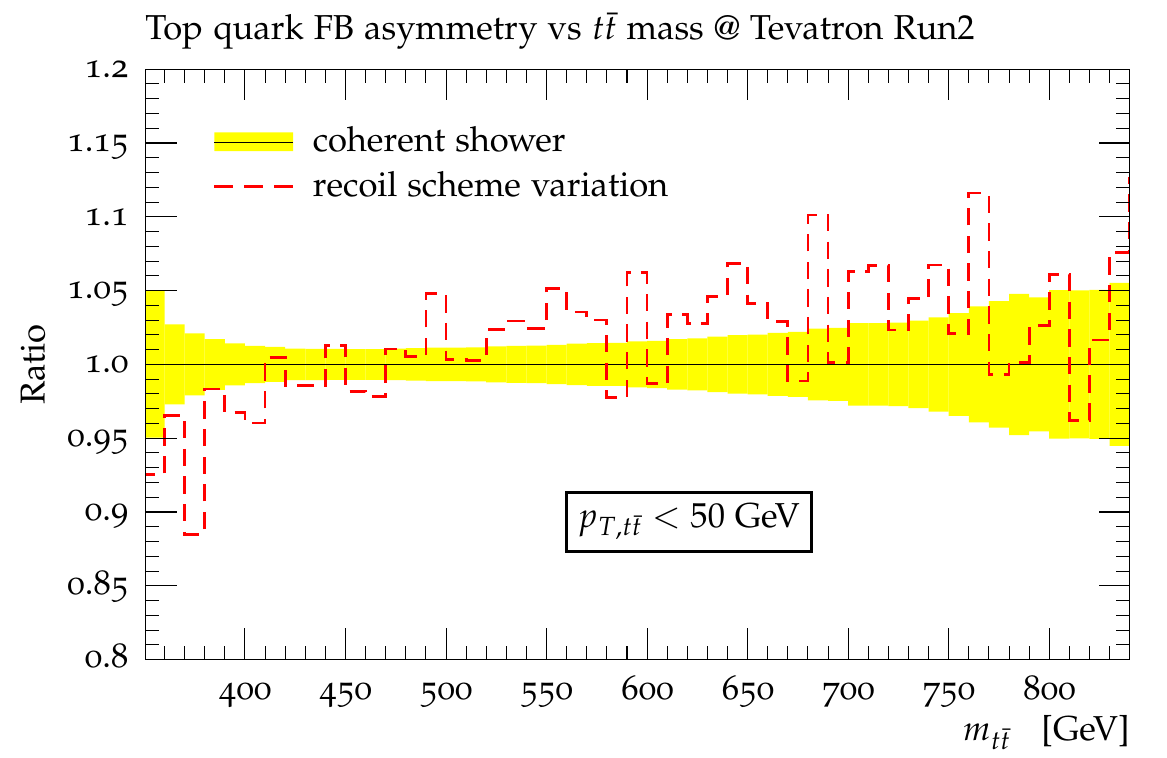}
  \includegraphics[width=0.33\textwidth,clip]{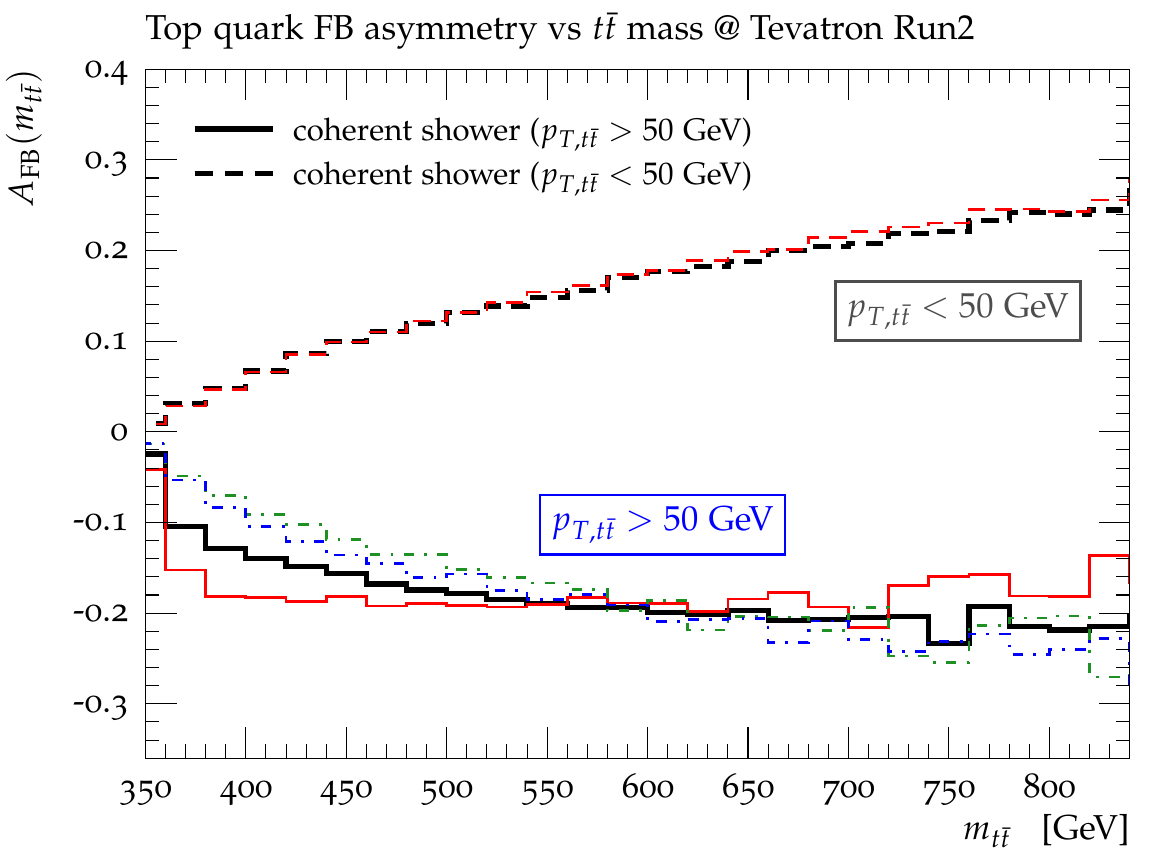}
  \includegraphics[width=0.33\textwidth,clip]{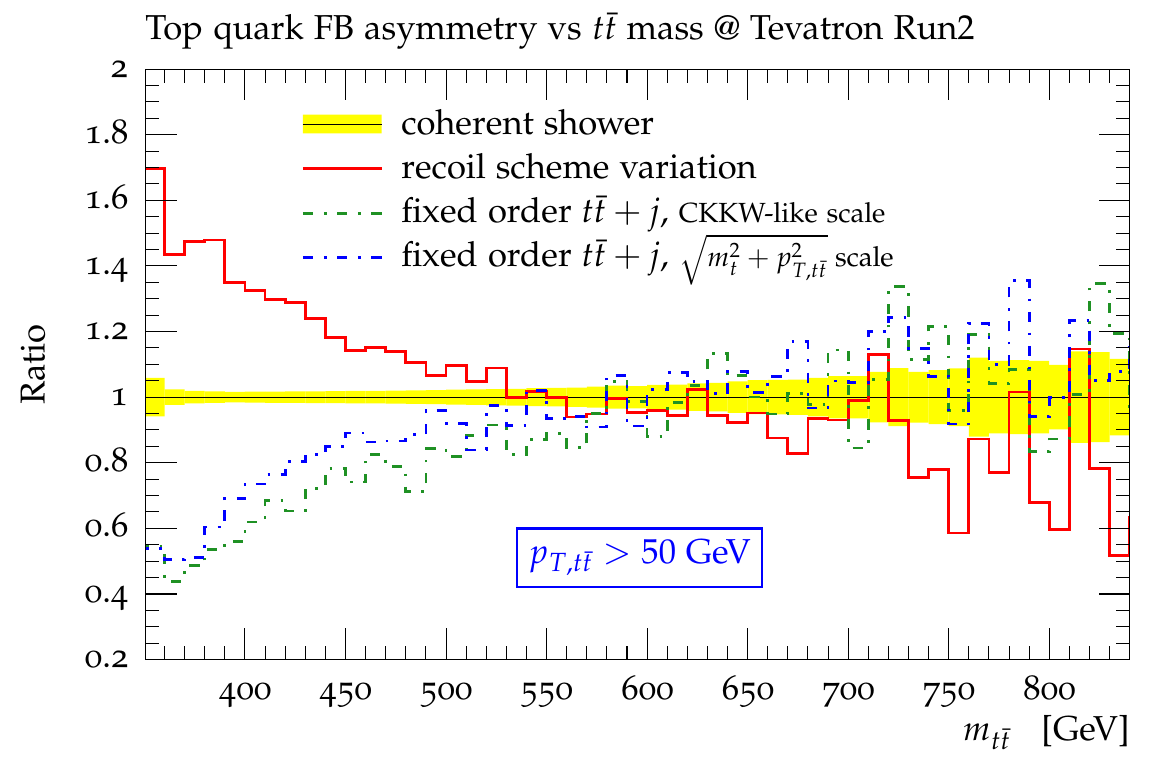}
  \caption{$A_\mathrm{FB}(m_{t\bar t})$ predictions (center) and ratio
    plots using colour coherent showering (here computed by
    \textsc{Sherpa}'s \textsc{CSshower}~\cite{Schumann:2007mg}) for a
    $p_{T,t\bar t}$ domain separation at $p_{T,t\bar t}=50\mathrm{\:GeV}$.
    A recoil scheme variation (as provided by \textsc{Sherpa} v.1.4.0)
    is also shown as well as reference curves for lowest-order $t\bar tj$\/
    production employing different scale choices.}
  \label{fig-cutpTtt}
  \vskip-2mm
\end{figure*}

By focusing on certain phase-space regions, one can amplify the
inclusive asymmetry, as shown in Tab.~\ref{tab-2}. The application of
cuts isolating large $m_{t\bar t}$ or low $p_{T,t\bar t}$ leads to an
increase of the positive asymmetry. The latter cut is particularly
useful to efficiently separate off the Sudakov region, i.e.\ the
domain of low $p_{T,t\bar t}$ that generates the positive contribution
to the overall asymmetry. Considering the anti-cut (real-emission)
region, one can test the most striking feature of the transition in
$p_{T,t\bar t}$ domains: that is the sign change in the asymmetry.
Because of the approximations discussed earlier, coherent showers
predict a stronger asymmetry flip at large $p_{T,t\bar t}$. For all
other cases, they however give enhancements similar to the NLO result.

\begin{figure}[t!]
  \centering\capstart
  \includegraphics[width=0.85\columnwidth,clip]{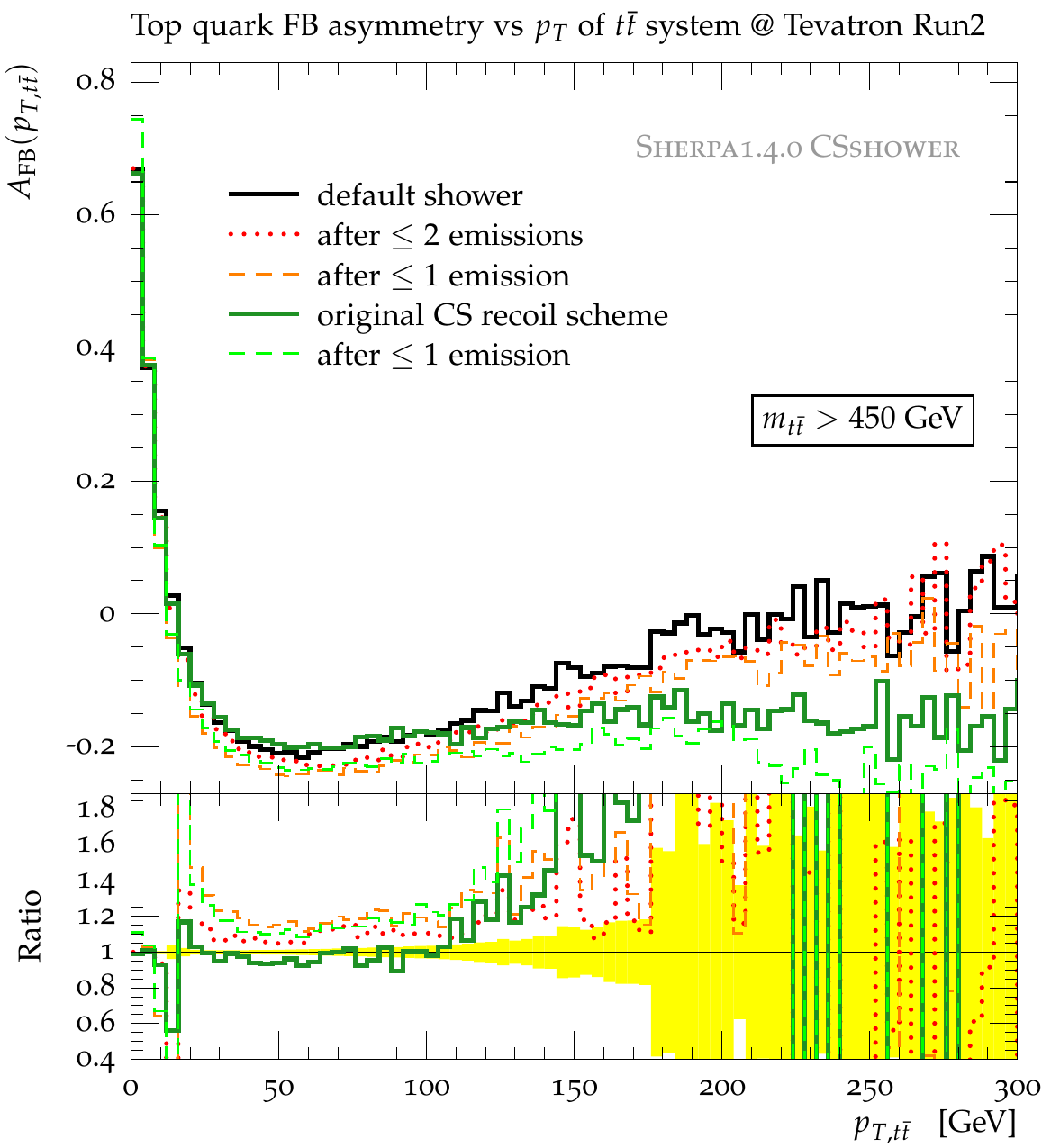}
  \caption{$A_\mathrm{FB}(p_{T,t\bar t})$ distributions obtained after
    coherent parton showering (computed with the
    \textsc{CSshower}~\cite{Schumann:2007mg}) for the constraint of
    large pair masses, $m_{t\bar t}>450\mathrm{\:GeV}$. The brighter
    dashed/dotted curves indicate the impact of the first two
    emissions. The green solid curve shows the outcome of a recoil
    scheme variation (as provided by \textsc{Sherpa}
    v.1.4.0~\cite{Gleisberg:2008ta}).}
  \label{fig-cutmtt}
\end{figure}

These single-cut, $A_\mathrm{FB}$ enhancing effects can also be found
at the differential level. We exemplify this in
Figs.~\ref{fig-cutpTtt}~and~\ref{fig-cutmtt} using \textsc{Sherpa}
\textsc{CSshower} results. Figure~\ref{fig-cutpTtt} is shown to
compare the different $A_\mathrm{FB}(m_{t\bar t})$ distributions
associated with the two $p_{T,t\bar t}$ domains. It exhibits the sign
flip in the asymmetry very clearly. The opposite scenario is presented
in Fig.~\ref{fig-cutmtt} where small top quark pair masses are
vetoed and $A_\mathrm{FB}$ is shown as a function of the pair
transverse momentum. As expected, the increase of $A_\mathrm{FB}$ is
most noticeable in the first few $p_{T,t\bar t}$ bins, those that are
close to zero. In both figures we also illustrate the impact of a
recoil scheme variation (as provided by \textsc{Sherpa} v.1.4.0):
differences only occur in the high $p_{T,t\bar t}$ region where the
amount of Tevatron data is too sparse to discriminate between the two
options.

\subsection{Top quark velocity dependence of the asymmetry}

\begin{figure}[t!]
  \centering\capstart
  \includegraphics[width=0.85\columnwidth,clip]{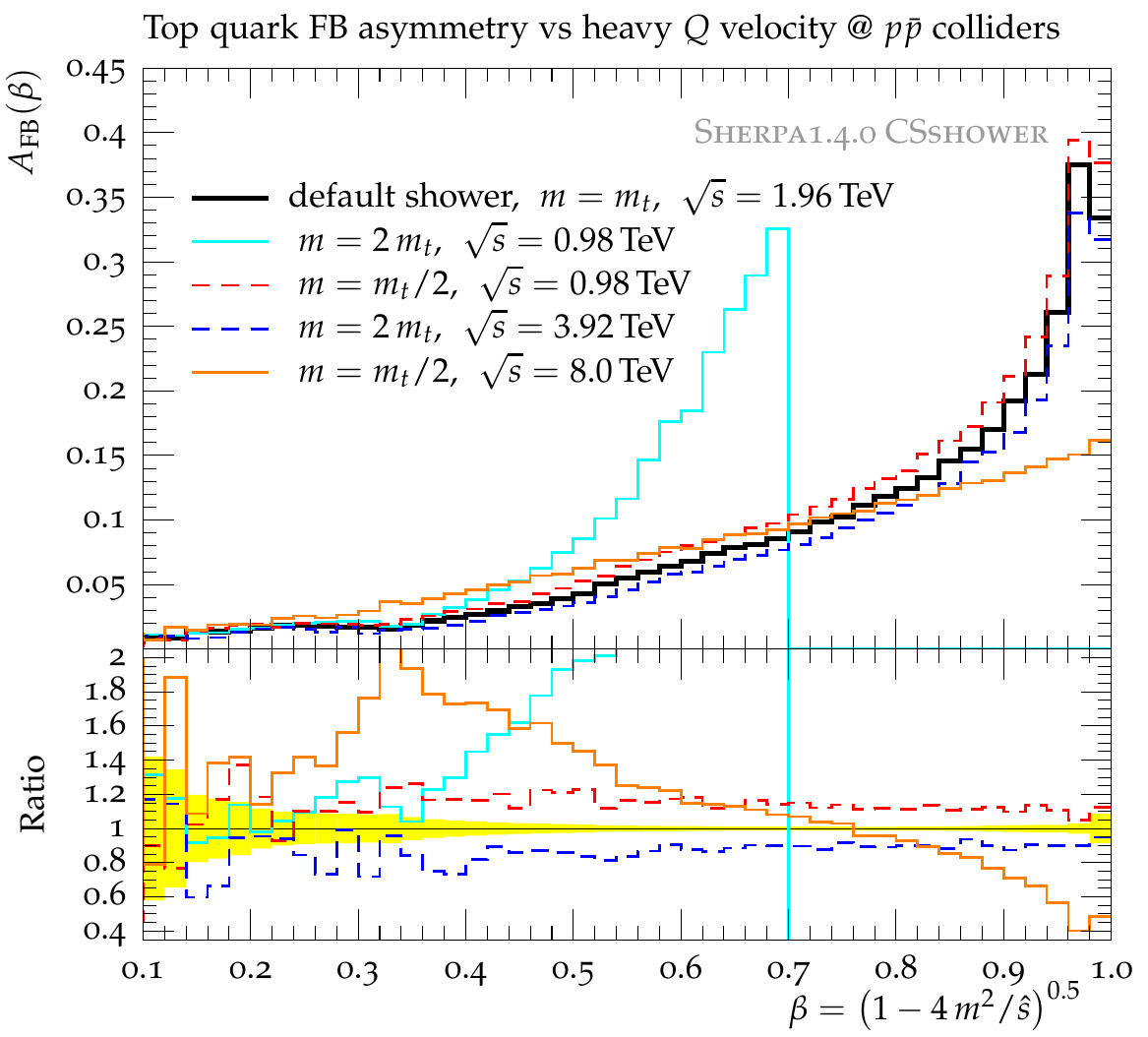}
  \caption{$A_\mathrm{FB}(\beta)$ distributions obtained from coherent
    parton showering (as implemented in the
    \textsc{CSshower}~\cite{Schumann:2007mg}) for varying ``top
    quark'' masses, $m$, and $p\bar p$\/ collider energies, $\sqrt s$.}
  \label{fig-beta}
\end{figure}

We want to briefly discuss how the coherence effect (generating
non-zero $A_\mathrm{FB}$) changes under variation of the top quark
mass and collider energy $\sqrt s$. It is thus convenient to analyze
the $\beta$\/ dependence of the asymmetry, which behaves roughly like
$A_\mathrm{FB}(m_{t\bar t})$. Figure~\ref{fig-beta} contains a number
of $A_\mathrm{FB}(\beta)$ curves obtained from coherent showering
using different values for the $m$\/ and $s$\/ parameters. Apart from
PDF effects, one finds, as expected, an approximately stable asymmetry
dependence, if mass and $\sqrt s$\/ are scaled equally. Investigating
the limiting cases, it is clear that the dramatic increase of
$A_\mathrm{FB}$ occurs as a consequence of reaching the collider
phase-space boundaries. The contribution to the inclusive asymmetry
however is small since the cross section of the associated phase-space
area is tiny.

\subsection{Top quark forward--backward asymmetry from the lepton perspective}

\begin{figure}[h!]
  \centering\vskip2pt\capstart
  \includegraphics[width=0.98\columnwidth,clip]{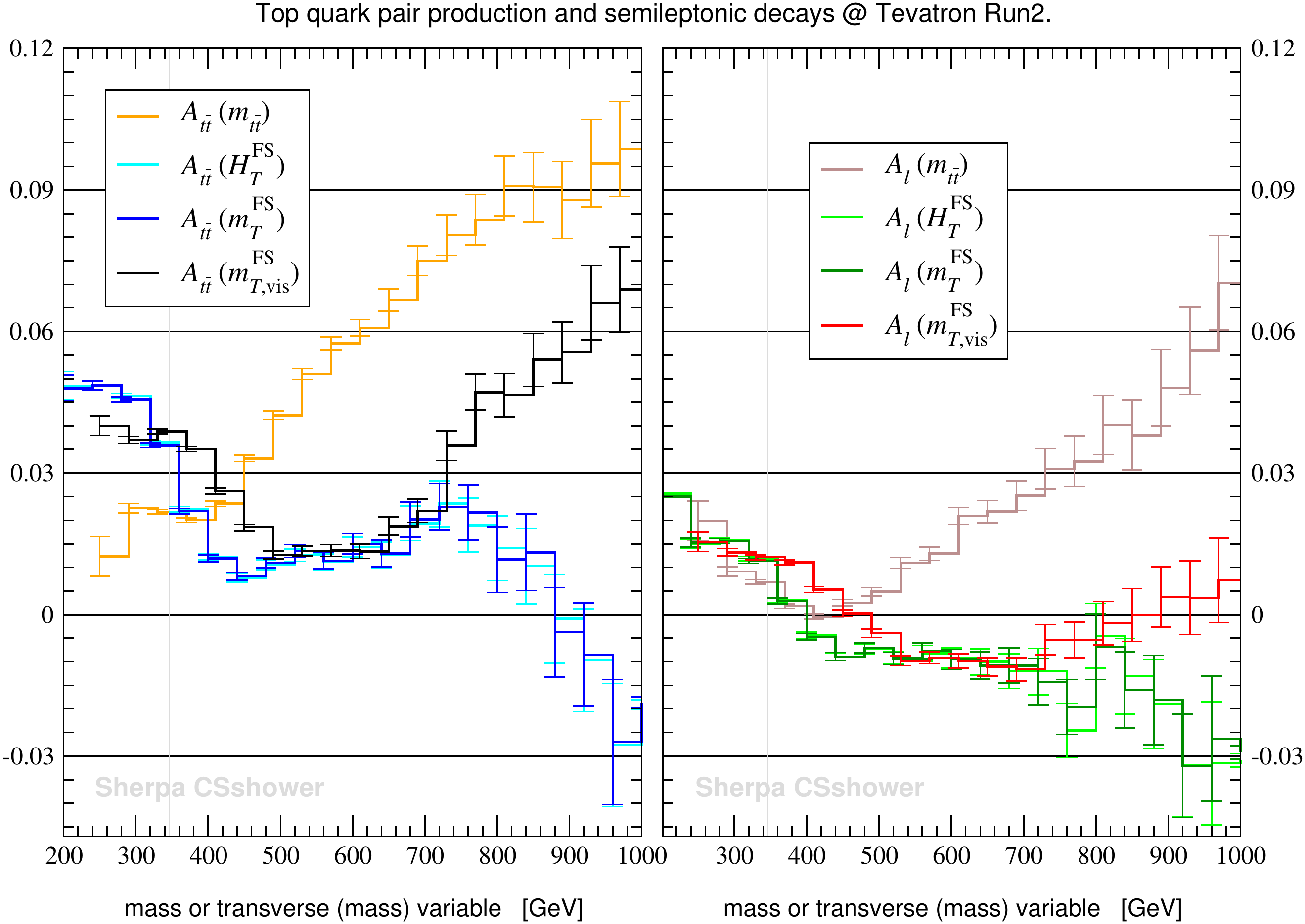}
  \caption{Differential forward--backward ($A_{t\bar t}\equiv A_\mathrm{FB}$)
    and lepton-based ($A_l$) asymmetries given in terms of various
    (transverse) mass observables~\cite{Falkowski:2012cu}. Here,
    \textsc{CSshower} results are shown.}
  \label{fig-alep}
  \vskip-4mm
\end{figure}

We gain information if we define the analysis not only in terms of the
forward--backward asymmetry, $A_{t\bar t}\equiv A_\mathrm{FB}$, but
also in terms of a lepton-based asymmetry $A_l$ (now including top
quark decays). This allows one to exploit the way these asymmetries
are correlated in the Standard Model and beyond; for the details see
Ref.~\cite{Falkowski:2012cu}. An example for coherent shower
predictions in terms of (transverse) mass observables, see also
\cite{Lykken:2011uv}, is given in Fig.~\ref{fig-alep} relating
results for reconstructed top quarks ($A_{t\bar t}$,
$A_{t\bar t}\,(m_{t\bar t})$, etc.) with results where this
reconstruction is not ($A_l$, $A_l\,(m^\mathrm{FS}_{T,\mathrm{vis}})$,
etc.) or only partially ($A_{t\bar t}\,(H^\mathrm{FS}_T)$,
$A_l\,(m_{t\bar t})$, etc.) required (the ``$O^\mathrm{FS}$''
superscript signifies that the transverse observables used here are
obtained from all final-state visible objects and $\slashed{E}_T$).
This introduces additional analysis handles and levels, and therefore
more directions to cross-check the data for consistency or potential
biases.

\section{Summary}\label{sec-sumup}

Even in the absence of a full NLO treatment, standard Monte Carlo
event generators can produce significant differential as well as
inclusive forward--backward asymmetries in top quark pair production.
These asymmetries arise from valid physics built into generators relying
on (approximately) coherent parton or dipole showering. While colour
coherent shower algorithms cannot be quantitatively correct in every
detail, they are able to capture important features as known from QCD
higher-order calculations. Based on these findings, coherent showers
are qualified to serve as fast tools in estimating the Standard Model
phenomenology of the forward--backward asymmetry.\footnote{For a
  recent example, see Ref.~\cite{Falkowski:2012cu}, where coherent
  shower predictions are used to estimate the correlation between
  $A_{t\bar t}\equiv A_\mathrm{FB}$ and $A_l$, the lepton-based
  asymmetry definable in the semileptonic decay channel.} However, if
used in data analyses to determine Monte Carlo corrections, one needs
to be aware of their previously unexpected asymmetry-generating
behaviour.

\acknowledgement

J.W. would like to thank the organizers and conveners of the HCP2012
symposium for arranging a fantastic conference. Additionally, financial
support by the local organizing committee is gratefully acknowledged.

\smallskip
We thank the \textsc{Rivet} \cite{Buckley:2010ar} and
\textsc{MCplots}\,\footnote{Visit the \texttt{mcplots.cern.ch}
  webpages for more information.}
software teams for providing these user-friendly interfaces for the
convenient visualization of MC results; in particular, we are grateful
to Anton Karneyeu for great help in producing the shower comparison
results shown here.

\newpage

\bibliography{jwinter_hcp+}\vskip-40mm
%
%
%
%

\end{document}